\documentclass[aps,prc,superscriptaddress,twoside,twocolumn,nofootinbib,showpacs]{revtex4-1}
\usepackage{amsmath,amssymb}
\usepackage{mathtools}
\usepackage{bbold}
\usepackage[usenames, dvipsnames]{xcolor}
\usepackage{braket}
\usepackage{graphicx,bm}
\usepackage{slashed}
\usepackage{booktabs}
\usepackage{tensor}
\usepackage{multirow}
\usepackage{mwe}
\usepackage{changes}
\usepackage{enumerate}

\allowdisplaybreaks

\usepackage{epsfig}
\usepackage{epstopdf}
\usepackage{subfigure}
\usepackage{float}

\newcommand{\pvec}[1]{\vec{#1}\mkern2mu\vphantom{#1}}

\allowdisplaybreaks

\begin{document}

\title{Constraining the nonanalytic terms in the isospin-asymmetry expansion of the \\ nuclear equation of state}

\title{Constraining the nonanalytic terms in the isospin-asymmetry expansion of the \\ nuclear equation of state}

\author{Pengsheng Wen}
\email{pswen2019@tamu.edu}
\affiliation{Cyclotron Institute, Texas A\&M University, College Station, TX 77843, USA}
\affiliation{Department of Physics and Astronomy, Texas A\&M University, College Station, TX 77843, USA}

\author{Jeremy W. \surname{Holt} }
\email{holt@physics.tamu.edu}
\affiliation{Cyclotron Institute, Texas A\&M University, College Station, TX 77843, USA}
\affiliation{Department of Physics and Astronomy, Texas A\&M University, College Station, TX 77843, USA}

\date{\today}

\begin{abstract}
We examine the properties of the isospin-asymmetry expansion of the nuclear equation of state 
from chiral two- and three-body forces. We focus on extracting the high-order symmetry energy 
coefficients that consist of both normal terms (occurring with even powers of the isospin asymmetry) 
as well as terms involving the logarithm of the isospin asymmetry that are formally nonanalytic around 
the expansion point of isospin-symmetric nuclear matter. These coefficients are extracted from 
numerically precise perturbation theory calculations of the equation of state coupled with a new set of
finite difference formulas that achieve stability by explicitly removing the effects of higher-order terms
in the expansion. We consider contributions to the symmetry energy coefficients from both two- and 
three-body interactions. It is found that the coefficients of the logarithmic terms are generically larger in 
magnitude than those of the normal terms from second-order perturbation theory diagrams, 
but overall the normal terms give larger contributions to the ground state energy. The high-order isospin-asymmetry terms are especially relevant at large densities where they effect the proton fraction in beta-equilibrium matter.
\end{abstract}

\maketitle

\section{Introduction}
The equation of state (EOS) of nuclear matter at arbitrary proton fraction and density is crucial
for understanding the structure and dynamics of neutron stars \cite{lattimer91,page92,steiner05,seif14}, 
the properties of neutron-rich nuclei \cite{horowitz01,chen05,centelles09,oyamatsu10,wang15},
and data from terrestrial heavy-ion collision experiments \cite{lynch09,horowitz14,mcintosh19}. 
For many years it has been assumed that the isospin-asymmetry 
expansion of the nuclear EOS at zero temperature follows a Maclaurin series:
\begin{align}
     \bar{E}(n,\delta) = A_0(n) + A_2(n)\delta^{2} + A_4(n)\delta^{4} + \cdots,
     \label{norms}
\end{align}
where $\bar{E}$ is the energy per particle, $n$ is the total baryon number density, and $\delta$ is the isospin asymmetry defined by $\delta=(n_n-n_p)/(n_n+n_p)$, where $n_n$ and $n_p$ are the number densities of 
neutrons and protons, respectively. Previous microscopic calculations 
\cite{PhysRevC.57.3488, PhysRevC.89.025806, LAGARIS1981470, MUTHER1987469} 
with realistic nuclear forces have 
found that the sum of the higher-order terms ($A_{2i\ge4}$) is relatively small around saturation density.
However, the convergence of the isospin-asymmetry expansion 
of the nuclear matter equation of state at higher densities is poorly known \cite{xu10}. 
Recent studies have shown that when the density is high, the terms $A_{2i\ge4}$ are significant 
for understanding neutron star structure. In particular the quartic term, and even the 
$\delta^6$ term, may enhance the proton fraction of $\beta$-equilibrium nuclear matter 
at high density \cite{cai12,seif14}, which can modify the critical density for the direct URCA 
process \cite{lattimer91} and lead to faster cooling of neutron stars \cite{steiner06}. Note that even when the density is 
low, the density dependence of $A_2$ is important for studying the collisions of neutron-rich 
nuclei \cite{PhysRevLett.78.1644}.

Recently, in Ref.\ \cite{kaiser15} it was discovered that 
perturbation theory calculations beyond the Hartree-Fock approximation can give rise to a 
modified form of the isospin-asymmetry expansion
\begin{align}
     & \hspace{-.3in} \bar{E}(n,\delta) = A_0(n) + A_2(n)\delta^2 \nonumber \\
     &+ \sum_{i > 1}\big ( A_{2i}(n) + A_{2i,l}(n) \log |\delta| \big )
     \delta^{2i},
     \label{logs}
\end{align}
which involves nonanalytic logarithm contributions that vanish
at both $\delta=0$ and $\delta=1$. The form in Eq.\ \eqref{logs} was obtained through analytic 
calculations carried out with a model zero-range contact potential \cite{kaiser15} and later 
verified numerically \cite{wellenhofer16} with a set of realistic chiral nuclear forces.
The modified form of the isospin-asymmetry expansion in Eq.\ \eqref{logs} has important
implications for the nuclear
symmetry energy $E_{\rm sym}(n)$, which is defined as the difference between the pure 
neutron matter equation of state and the isospin-symmetric nuclear matter equation of state 
at a given density: $E_{\rm sym}(n) = \bar{E}(n,\delta=1) - \bar{E}(n,\delta=0)$. We note that for 
both expansions in Eqs.\ \eqref{norms} and \eqref{logs}, the symmetry energy is given as the sum 
$E_{\rm sym}(n) = \sum_{i > 0}A_{2i}(n)$. The symmetry energy therefore contains no information
about the strength of the logarithmic contributions, which affect the ground state energy of nuclear matter at intermediate values of the isospin asymmetry $\delta$. This could be especially
important in the context of neutron stars, since the $\delta^4\ln |\delta|$ contribution peaks at
a value of $\delta = e^{-1/4} \simeq 0.78$, and the higher-order terms peak at even larger values of
$\delta$ close to that found in neutron stars. Nevertheless, recent studies \cite{wellenhofer16,somasundaram20} have suggested that around saturation density the regular terms of the isospin-asymmetry expansion of the nuclear EOS are dominant over the logarithmic terms.

In the present work we extract for the first time the high-order terms of the isospin-asymmetry expansion of the nuclear EOS up to sixth-order in powers of $\delta$. We employ chiral effective field theory (EFT) nuclear interactions including two-body and three-body forces.
In chiral EFT, the nuclear 
interactions are organized in a systematic expansion of the ratio $(Q/\Lambda_\chi)^v$, where $Q$ is the characteristic energy scale of the system (or the pion mass) and $\Lambda_\chi \sim 1$\,GeV is the chiral symmetry breaking scale. The interactions in chiral EFT naturally include short-, intermediate-, and long-range interactions \cite{epelbaum09rmp,MachleidtChiral2011}. The short-range two-body interactions are fitted to nucleon-nucleon (NN) data, such as nucleon-nucleon scattering phase shifts and properties of the deuteron. The three-nucleon forces (3NF) that appear at N2LO (next-to-next-to-leading order) in chiral EFT are essential for microscopic descriptions of nuclei \cite{holt20,PhysRevC.87.034307,navratil16,doi:10.1146/annurev-nucl-102313-025446,Holt:2013fwa,Whitehead:2018bfs} and the properties of nuclear matter in neutron stars \cite{baldo1997microscopic,BOGNER200559,PhysRevLett.85.944,Kaiser2012,PhysRevC.90.054322,hebeler10,Gandolfi:2011xu,Coraggio:2012ca,Gezerlis:2013ipa,Holt:2014hma}. For the chiral nuclear forces employed in the present work, the low-energy constants associated with short-range three-body interactions are fitted to the binding energies of $^3$H and $^3$He as well as the beta-decay lifetime of $^3$H \cite{gardestig06,gazit09,PhysRevC.91.054311}.

To extract the high-order isospin-asymmetry coefficients, we employ a modified finite difference method 
extended to account for both the normal and divergent logarithm terms,
$A_{2i}$ and $A_{2i;l}$. In addition, the method explicitly eliminates the influence of higher-order 
terms, leading to enhanced convergence. By varying the step size in the normal finite difference 
method, we can build a chain of 
equations to solve for the coefficients $A_{2i}$ and $A_{2i;l}$ numerically. After applying this method,
we find that the second-order term $A_{2}\delta^2$ remains the dominant component of the 
asymmetric nuclear matter equation of state, but the inclusion of the logarithmic terms increases 
the precision of the isospin-asymmetry expansion.

As an application, we consider the proton fraction in $\beta$-equilibrium nuclear matter, 
which can be determined be enforcing chemical equilibrium for the weak reactions
\begin{align}
	n \to p + e^- + \bar{\nu}_e, \quad p + e^- \to n + \nu_e
	\label{eq:beta_proc}
\end{align}
that comprise the direct URCA process \cite{lattimer91}. 
The equilibrium of the above two processes is closely related to the symmetry energy, 
and by expanding the EOS up to $6^{\rm th}$-order in powers of $\delta$ we extract the associated proton fraction. We find that for all chiral potentials employed in this work, the $4^{\rm th}$-order terms reduce the proton fraction at low-densities $n<1.5n_0$, where $n_0=0.16$\,fm$^{-3}$ is the saturation density of nuclear matter. At higher densities, however, there is greater model dependence on the choice of chiral interaction. Since the regular terms receive contributions at both the Hartree-Fock level and 
second-order perturbation theory, 
they play a more important role in the proton fraction than the logarithmic terms.

The paper is organized as follows. In Section \ref{formalism} we briefly 
review the formalism for computing the nuclear matter equation of state in many-body perturbation 
theory. In Section \ref{sec:fdiff} the normal and modified finite-difference methods are described 
in detail. In section \ref{results} we extract the numerical results for the coefficients $A_{2}$, $A_{2i}$ 
and $A_{2i,l}$ from Hartree-Fock and second-order perturbation theory calculations of the 
asymmetric nuclear matter equation of state obtained from the modified finite difference method. 
We end with a summary and discussion.

\section{EOS from many-body perturbation theory}
\label{formalism}

To calculate the ground state energy of nuclear matter, we start with the Hamiltonian
\begin{align}
     H = \sum_{i}\frac{p_i^2}{2M} + \frac{1}{2}\sum_{ij}V_{ij} + \frac{1}{6}\sum_{ijk}V_{ijk}
     = H_0 + H_I,
\end{align}
where $\vec{p}_i$ is the momentum of nucleon $i$, $V_{ij}$ is the two-body interaction between 
nucleon $i$ and $j$, and $V_{ijk}$ is the three-body interaction between nucleons $i$, $j$ and $k$. 
The perturbing interaction is defined by $H_I = \frac{1}{2}\sum_{ij}V_{ij} + \frac{1}{6}\sum_{ijk}V_{ijk}$.

At the Hartree-Fock level
\begin{align}\label{eq:eos_HF_ori}
     E = \langle \Phi_0 | H_I |\Phi_0\rangle,
\end{align}
where $|\Phi_0\rangle$ is the noninteracting ground state and $H_I$ is the interaction part of 
nuclear Hamiltonian. The two-nucleon force contribution reads
\begin{align}
     E^{(1)}_{NN} = \frac{1}{2}\sum_{ij}\langle ij|V_{ij} | ij\rangle,
\end{align}
where $i,j$ refer to filled states in the noninteracting Fermi sea. 
For homogeneous nuclear matter at zero temperature in the thermodynamic limit
\begin{align}
     E_{NN}^{(1)} =& \frac{1}{2}\sum_{s_1,s_2}\sum_{t_1,t_2} 
     \left[\frac{\Omega}{(2\pi)}\right]^3\int d^3 p_i d^3 p_j
	  \langle x_{ij} | 
     V_{ij} 
	  | x_{ij}\rangle\nonumber\\
     &\times\Theta(k_F^i - |\vec{p}_i|)\Theta(k_F^j - |\vec{p}_j|),
\end{align}
where $\Omega$ is the volume, $k_F^{i}$ is the Fermi momentum of particle $i$, and
\begin{align}
	|x_{ij}\rangle  = | s_i,t_i,\pvec{p}_i;s_j,t_j,\pvec{p}_j\rangle.
\end{align}
The values of $k_F^{i, j}$ depend on the Fermi momentum of symmetric nuclear matter $k_F$ 
and the isospin-asymmetry parameter $\delta$. 
For neutrons we have $k_F^n = k_F(1+\delta)^{1/3}$, while for protons we have $k_F^p = k_F(1-\delta)^{1/3}$. 
Similarly, for three-body interactions the Hartree-Fock contribution to the ground state energy of 
nuclear matter is 
\begin{align}
     E_{3NF}^{(1)} =& \frac{1}{6}\sum_{ijk}
     \langle ijk | V_{ijk} | ijk\rangle\nonumber\\
     =& \frac{1}{6}\left[\frac{\Omega}{(2\pi)^3}\right]^3
     \sum_{s_i,s_j,s_k}\sum_{t_i,t_j,t_k}\int d^3 p_1 d^3 p_2 d^3 p_3\nonumber\\
	  &\times \langle x_{ijk}|
     V_{ijk}|
	  x_{i j k}\rangle\nonumber\\
     &\times \Theta(k_F^i - |\vec{p}_i|)\Theta(k_F^j - |\vec{p}_j|)\Theta(k_F^k - |\vec{p}_k|),
\end{align}
where
\begin{align}
	|x_{i,j,k}\rangle = |s_i,t_i,\pvec{p}_i; s_j,t_j,\pvec{p}_j;s_k,t_k,\pvec{p}_k\rangle
\end{align}
is the state specified by the spin, isospin, and momentum of particles $i$, $j$ and $k$.

The energy per particle at the Hartree-Fock level can be expanded with respect to $\delta$ 
in a Maclaurin series \cite{wellenhofer16},
\begin{align}
    \frac{E^{(1)}}{A} = A_0 + A_2 \delta^2 + \mathcal{O}(\delta^{4})+\cdots,
\end{align}
where $A = \rho V$ is the number of particles and $\rho={2k_f^3}/{3\pi^2}$ is the nucleon number density. The value of these parameters $A_{2i}$ can be extracted by the normal finite difference method described in the next section.

The second-order perturbative contribution to the ground-state energy is given by
\begin{align}
     E^{(2)} = \left\langle \Phi_0 \left|
     \frac{H_I}{E_0 - H_0}
     \left(1-| \Phi_0\rangle\langle\Phi_0|\right)H_I
     \right|{\Phi_0}\right\rangle,
\end{align}
where $E_0$ is the energy of the noninteracting ground state. The contribution from two-body forces alone reads
\begin{align}
     E^{(2)}_{NN} =& 
     -\frac{1}{4}\sum_{s_i,s_j,s_l,s_m}\sum_{t_i,t_j,t_l,t_m}
     \int d^3 p_i d^3 p_j d^3 p_l d^3 p_m\nonumber\\
	  &\times\frac{\langle x_{i j}| 
     V_{ij} 
	  | x_{lm}\rangle^2}
     {{E_i + E_j} - E_l - E_m}\Theta(k_F^l - |\vec{p}_l|)\nonumber\\
     &\times \Theta(k_F^m - |\vec{p}_m|)
     \Theta(|\vec{p}_i| - k_F^i )\Theta(|\vec{p}_j| - k_F^j),
\end{align} 
where $E_i = p_i^2/2M^2$ is the kinetic energy of particle $i$ assuming a free spectrum. The contribution to the ground-state energy from 3-body interactions at second order in perturbation theory can be conveniently approximated using the two-body normal-ordering approximation \cite{holt20}. Below we will discuss specific features of this approximation as it relates to the extraction 
of higher-order symmetry energy coefficients.

\section{Finite difference method}
\label{sec:fdiff}

In this section, we describe the normal and modified finite difference methods. 
The normal finite difference method can be used to extract the coefficients in a Maclaurin or Taylor series. 
We note if a function $f(\delta)$ can be expanded according to a Maclaurin series of the form 
$f(\delta)=\sum A_{2i}\delta^{2i}$, 
then the $2i^{\rm th}$-order derivative of this function at position $\delta=0$ is $\delta$-independent 
and proportional to the coefficient $A_{2i}$. 
However, the presence of a term such as $\log|\delta|$ in Eq.~\eqref{logs} breaks the Maclaurin series 
and makes the coefficient $A_{2i}+A_{2i;l}\log|\delta|$ explicitly $\delta$-dependent.

\subsection{Normal finite difference method}
The original finite difference method can be used to extract the coefficients in a Maclaurin series
\begin{align}\label{eq:eos_exp_hf}
     \bar{E} = A_0 + \sum_{i=1}^{\infty} A_{2i} \delta^{2i}
     = A_0 + \sum_{i=1}^\infty A_{2i} \eta^i,
\end{align}
where $\eta = \delta^2$. 
The finite difference method uses a certain number of points $\mathcal{N}$ 
and uniform grid space $\Delta\delta$ to calculate the numerical value of the derivative of a function 
for a given accuracy $\mathcal{O}(\Delta \delta^{\mathcal{A}})$. 
Derivatives with respect to the variable $\eta$ in $\bar{E}$ can be obtained from 
the forward finite difference:
\begin{align}
     \frac{\partial^{i} \bar{E}}{\partial \eta^{i}} \approx
     \frac{1}{\Delta\delta^{2i}}\sum_{m=0}^{\mathcal{N}(\mathcal{A})}
     \omega_{m}^{i;\mathcal{A}}\bar{E}(\sqrt{m}\Delta\delta) 
     + \mathcal{O}(\Delta\delta^\mathcal{A}).
\end{align}
The number of points $\mathcal{N}$ and the coefficients $\omega_{m}^{i;\mathcal{A}}$ are determined by 
the order of derivative $2i$ and the accuracy $\mathcal{A}$.
Specific values for the coefficients can be found in Ref.\ \cite{10.2307/2008770}. 
Therefore, the coefficient $A_{2i}$ in the Maclaurin series can be calculated numerically by
\begin{align}\label{eq:MFDM}
     A_{2i} \approx 
     A_{2i}^{\mathcal{A};\Delta\delta} = 
     \frac{1}{i!\Delta\delta^{2i}}\sum_{m=0}^{\mathcal{N}(\mathcal{A})}
     \omega_{m}^{i;\mathcal{A}}\bar{E}(\sqrt{m}\Delta\delta) 
     + \mathcal{O}(\Delta\delta^\mathcal{A}).
\end{align}
The normal finite difference method is suitable when there are no divergent $\log|\delta|$ terms 
in a Maclaurin series. 
We can always choose a high accuracy to remove the influence from higher-order terms 
in the numerical value $A_{2i}^{\mathcal{A};\Delta\delta}$. 
For example, if the accuracy is $\mathcal{A} = 4$, 
there will not be terms proportional to $\Delta\delta^{0<i<4}$ 
in the expression of $A_{2i}^{\mathcal{A};\Delta\delta}$, 
and the influence from high-order powers of $\Delta\delta$ can be ignored when a small uniform grid space is applied. 
However, we will see that when there are log terms in the expansion of the EOS, 
we cannot remove the influence from high-order $\Delta\delta$ terms by simply increasing the accuracy.

\begin{figure}[t]
     \centering
     \includegraphics[scale=0.5]{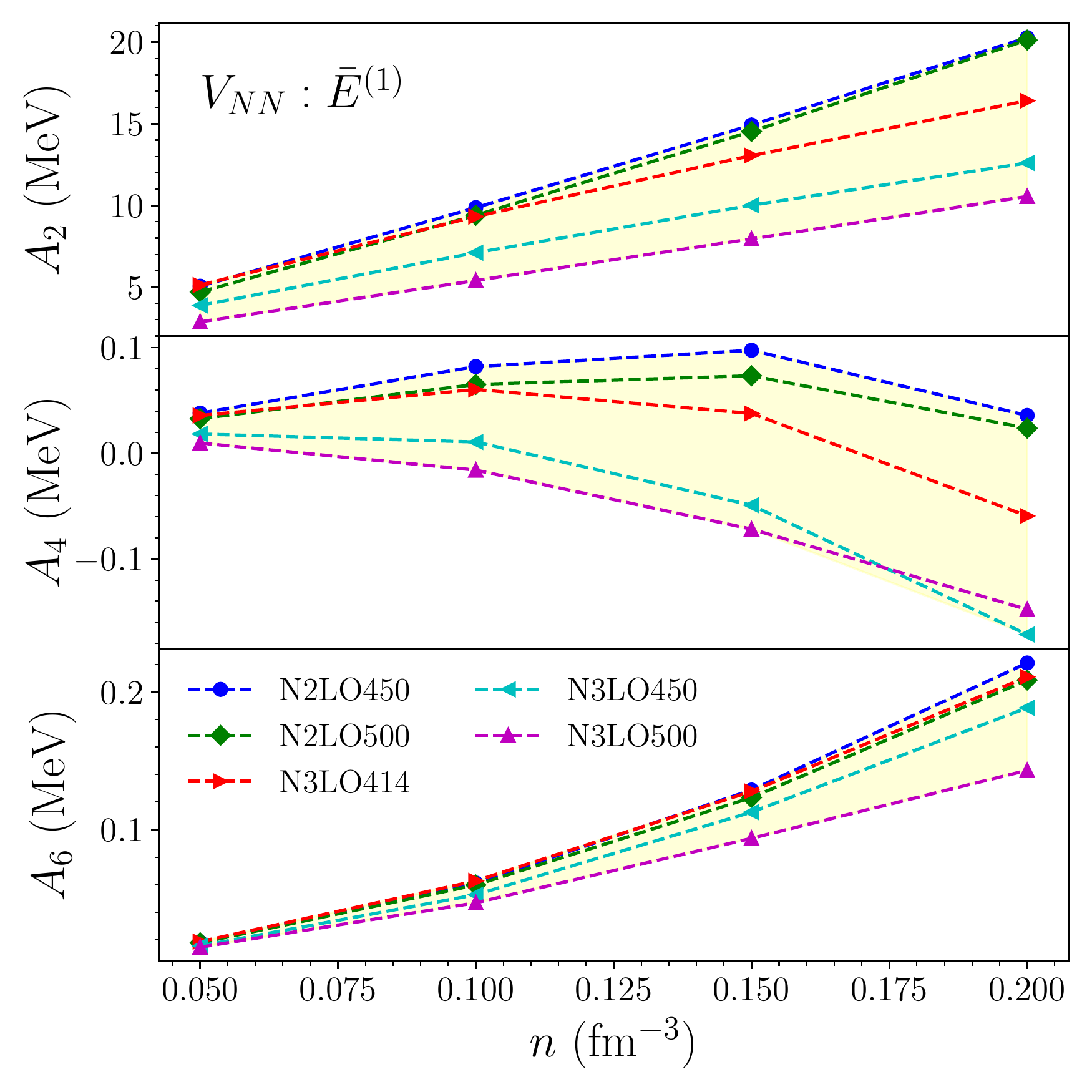}
     \caption{Density dependence of the $A_2$, $A_4$, and $A_6$ Maclaurin series coefficients of the isospin asymmetry expansion of the nuclear EOS for various chiral NN interactions at the Hartree-Fock level.}
     \label{fig:As_2NF_HF}
\end{figure}

\subsection{Modified finite difference method}
The isospin-asymmetry dependence of the EOS at second order in perturbation theory can be expanded according to
\begin{align}
     \bar{E} = A_0 + A_2\eta + 
     \sum_{i=2}^{\infty}\left(A_{2i}+\frac{1}{2}A_{2i;l}\log|\eta|\right)\eta^i.
\end{align}
This expression contains logarithmic terms for which the normal finite difference method fails to 
extract coefficients at a given order of $\delta$ free from the influence of higher orders. 
To illustrate this point, we show the formulas for 
$A_{4}^{\mathcal{A};\Delta\delta}$ and $A_{6}^{\mathcal{A};\Delta\delta}$ 
produced by the normal finite difference method when the accuracy is $\mathcal{A}\ge2$:
\begin{align}
	  \hspace{-.3in}A_{4}^{\mathcal{A};\Delta\delta} &=
     \frac{1}{2\Delta\delta^4}\sum_{m = 0}^{\mathcal{N}(\mathcal{A})}
     \omega_{m}^{2;\mathcal{A}}
     \bar{E}(\sqrt{m}\Delta\delta) \\ \nonumber
	  &\hspace{-.55in}= A_{4}+A_{4;l}\log\Delta\delta
	  +C_1^{4;\mathcal{A}}A_{4;l}+C_2^{4;\mathcal{A}}A_{6;l}\Delta\delta^2
     +\mathcal{O}(\Delta\delta^{4}),\\
	  A_{6}^{\mathcal{A};\Delta\delta} = &
     \frac{1}{6\Delta\delta^6}\sum_{m=0}^{\mathcal{N}(\mathcal{A})}
     \omega_{m}^{3;\mathcal{A}}
     \bar{E}(\sqrt{m}\Delta\delta)\\
	  =&A_{6}+A_{6;l}\log\Delta\delta\nonumber\\ \nonumber
	  +&C_1^{6;\mathcal{A}}A_{6;l}+\frac{C_2^{6;\mathcal{A}}A_{4;l}}{\Delta\delta^2}
     +C_3^{6;\mathcal{A}}A_{8;l}\Delta\delta^2
     +\mathcal{O}(\Delta\delta^4),
\end{align}
where the coefficients $C_i$ depend on the accuracy $\mathcal{A}$. 
Note that the $\mathcal{O}(\Delta\delta^2)$ terms, $A_{6;l}\Delta\delta^2$ and $A_{8;l}\Delta\delta^2$, 
influence the numerical results for $A_{4}^{\mathcal{A};\Delta\delta}$ and $A_{6}^{\mathcal{A};\Delta\delta}$, 
respectively, even though a high accuracy is applied. The appearance of these terms is due to 
the presence of the logarithmic term at the next order and cannot be removed by 
the normal finite difference method. 
We now develop an extended finite difference method to remove them.

\begin{figure}[t]
     \centering
     \includegraphics[scale=0.56]{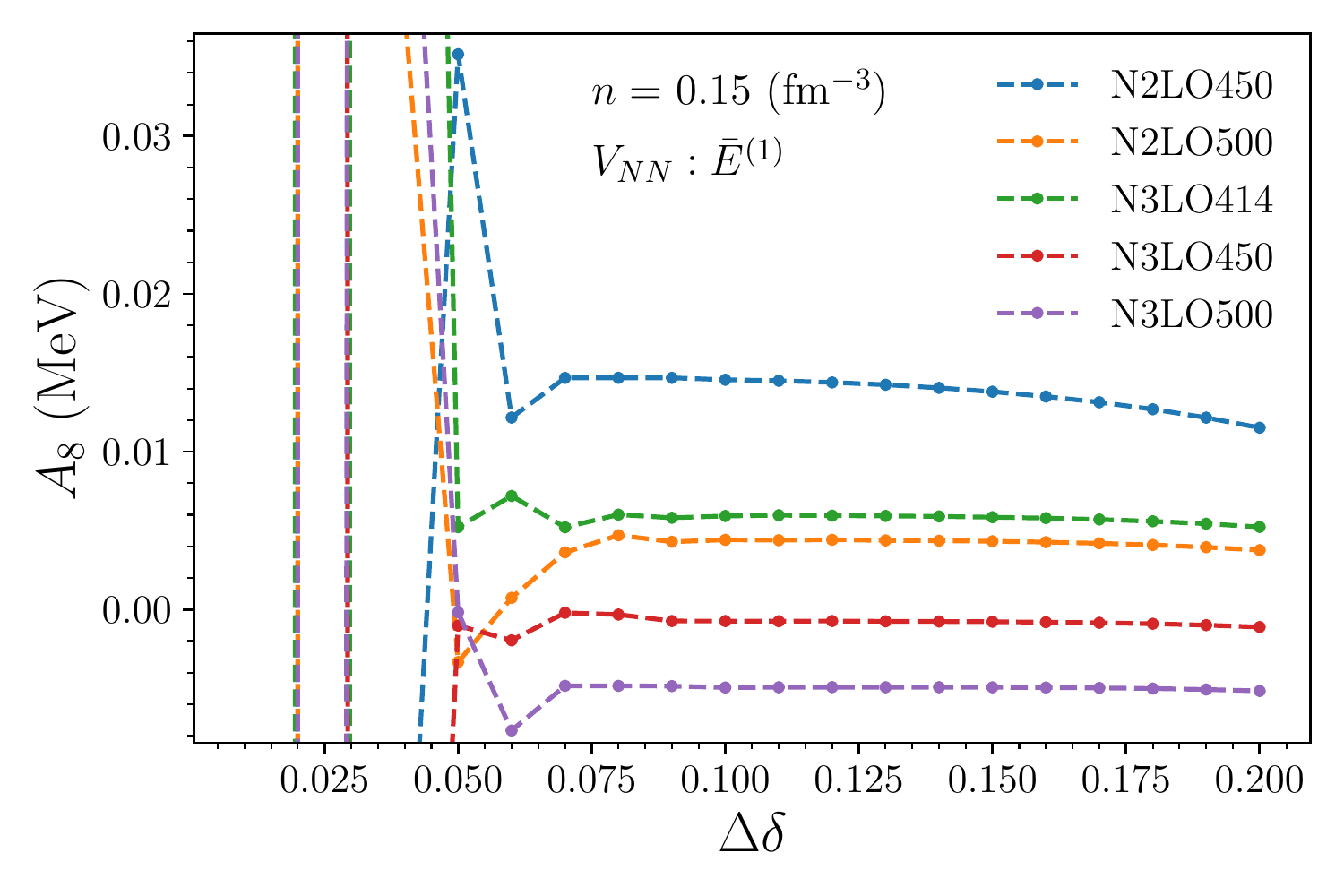}
     \caption{The $A_8$ Maclaurin series coefficient in the isospin asymmetry expansion of the nuclear EOS at density $n=0.15$\,fm$^{-3}$ calculated in the Hartree-Fock approximation from various NN forces in chiral EFT. The plateaus in $\Delta \delta$ indicate the extracted value of the coefficient.}
     \label{fig:A8_2NF_HF}
\end{figure}

The values of $A_{4}$, $A_{4;l}$, $A_{6}$, and $A_{6;l}$ can be deduced by using a set of accuracies 
that will give us a chain of equations to solve \{$A_{4}$, $A_{4;l}$, $A_{6}$, $A_{6;l}$\} as unknown factors 
and to remove other terms from the next order. We choose three different accuracies 
${\mathcal{A}_1, \mathcal{A}_2, \mathcal{A}_3}$ to calculate $A_4^{\mathcal{A};\Delta\delta}$:
\begin{align}
    \label{mfd1}
     A_{4}^{\mathcal{A}_i;\Delta\delta} 
	  \approx& A_{4}+A_{4;l}\log\Delta\delta\\ \nonumber
	  &+C_1^{4;\mathcal{A}_i}A_{4;l}+C_2^{4;\mathcal{A}_i}A_{6;l}\Delta\delta^2, \quad i = 1,2,3.
\end{align}
The differences between the three $A_4^{\mathcal{A}_i;\Delta\delta}$ will generate
\begin{align}
     \mathcal{D}_{1-2} = 
     \frac{A_4^{\mathcal{A}_1;\Delta\delta}-A_4^{\mathcal{A}_2;\Delta\delta}}
     {C_1^{4;\mathcal{A}_1}-C_1^{4;\mathcal{A}_2}}\approx A_{4;l} 
     + K_1^\mathcal{D} A_{6;l}\Delta\delta^2,\nonumber\\
     \mathcal{D}_{2-3} = 
     \frac{A_4^{\mathcal{A}_2;\Delta\delta}-A_4^{\mathcal{A}_3;\Delta\delta}}
     {C_1^{4;\mathcal{A}_2}-C_1^{4;\mathcal{A}_3}}
     \approx A_{4;l} + K_2^\mathcal{D} A_{6;l}\Delta\delta^2,
\end{align}
where $K^\mathcal{D}_{1(2)}$ are constants.  
By using $\mathcal{D}_{1-2}$ and $\mathcal{D}_{2-3}$, we can derive
\begin{align}\label{eq:mfdm_A4l}
     A_{4;l}\approx \left[
     \frac{\mathcal{D}_{1-2}}{K_1^\mathcal{D}}
     -\frac{\mathcal{D}_{2-3}}{K_2^\mathcal{D}}
     \right]\frac{K_1^\mathcal{D}K_2^\mathcal{D}}{K_2^\mathcal{D}-K_1^\mathcal{D}}.
\end{align}

\begin{table*}[]
\begin{tabular}{|l|l|l|l|l|l|l|l|l|l|l|l|l|l|}
\hline
$C_1^{4;2}$ & $C_1^{4;3}$ & $C_1^{4;4}$ & $C_2^{4;2}$ & $C_2^{4;3}$ & $C_2^{4;4}$ &
$K_1^{\mathcal{D}}$ & $K_2^{\mathcal{D}}$ &
$K_1^{\mathcal{R};1-2}$ & $K_2^{\mathcal{R};1-2}$ & $K_3^{\mathcal{R};1-2}$ &
$K_1^{\mathcal{R};2-3}$ & $K_2^{\mathcal{R};2-3}$ & $K_3^{\mathcal{R};2-3}$\\ \hline
    $0.3007$ & $0.1325$ & $0.0322$ & $-1.8705$ & $-1.1041$ & $-0.8415$ &
$-4.5568$ & $-2.6181$ &
$0.3710$ & $-0.0407$ & $0.3710$ &
$0.2827$ & $-0.0818$ & $0.2827$\\ \hline
\end{tabular}
\caption{The values of the coefficients in the modified finite difference method of Eqs.\ \eqref{mfd1}-\eqref{eq:mfdm_A4r}. The set of accuracies employed is $\{\mathcal{A}_1,\mathcal{A}_2,\mathcal{A}_3\}=\{2,3,4\}$.}
\label{table:MFDM}
\end{table*}

\begin{figure*}[t]
     \centering
     \includegraphics[scale=0.4]{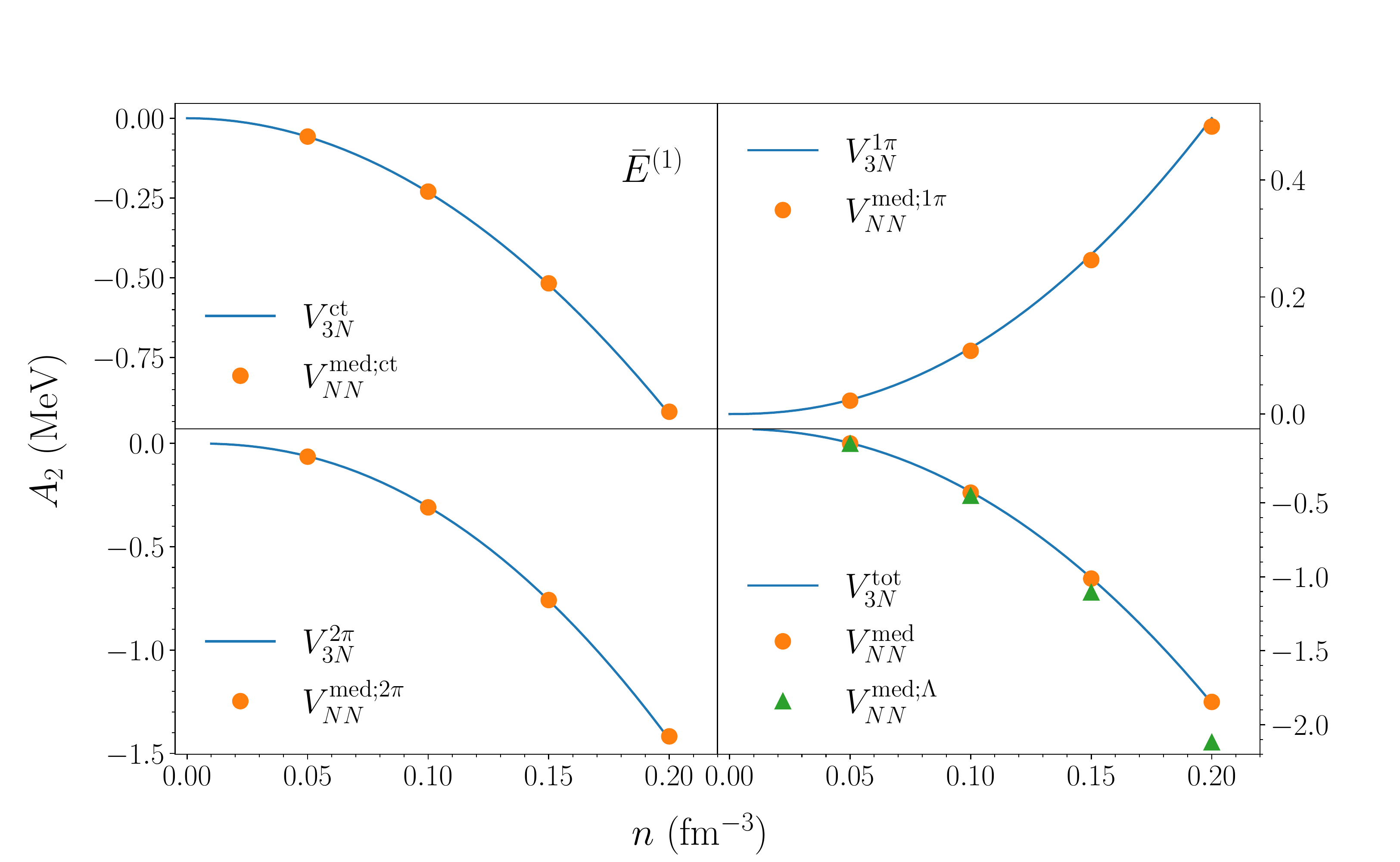}
     \caption{Density dependence of the $A_2$ Maclaurin series coefficient in the isospin-asymmetry expansion of the nuclear EOS at the Hartree-Fock level from 3N forces. Exact $V_{3N}$ (blue lines) and approximate $V_{NN}^{\rm med}$ (orange dots) contributions to $A_2$ from individual 3NF topologies (contact, $1\pi$ exchange, and $2\pi$ exchange) are shown. In the bottom-right panel, we show the sum of the three contributions and also the effect (green triangles) of including the multiplicative regulating function in Eq.\ \eqref{2Nreg}.}
	  \label{fig:A2_3NF_HF}
\end{figure*}

The idea to extract $A_{4}$ is similar. We can make $A_{6;l}$ terms vanish at first by,
\begin{align}
	\mathcal{R}_{1-2} =& 
     \frac{A_4^{\mathcal{A}_1;\Delta\delta}}{C_{2}^{4;\mathcal{A}_1;\Delta\delta}}-
     \frac{A_4^{\mathcal{A}_2}}{C_{2}^{4;\mathcal{A}_2}}\nonumber\\
     \approx&K^{\mathcal{R};1-2}_{1}A_{4;l}+K^{\mathcal{R};1-2}_2 A_{4;l}\log\Delta\delta
     +K^{\mathcal{R};1-2}_{3}A_{4}\nonumber\\
     =&F^{\mathcal{R};1-2}[\Delta\delta]+K^{\mathcal{R};1-2}_3A_{4}\\
	  \mathcal{R}_{2-3} =&
     \frac{A_4^{\mathcal{A}_2;\Delta\delta}}{C_{2}^{4;\mathcal{A}_2}}-
     \frac{A_4^{\mathcal{A}_3;\Delta\delta}}{C_{2}^{4;\mathcal{A}_3}}\nonumber\\
     \approx&K^{\mathcal{R};2-3}_{1}A_{4;l}+K^{\mathcal{R};2-3}_2 A_{4;l}\log\Delta\delta
     +K^{\mathcal{R};2-3}_{3}A_{4}\nonumber\\
     =&F^{\mathcal{R};2-3}[\Delta\delta]+K^{\mathcal{R};2-3}_3A_{4}
\end{align}
Therefore, the $A_{4}$ can be extracted by,
\begin{align}
	A_{4} &\approx \left(\frac{\mathcal{R}_{1-2}}{F^{\mathcal{R};1-2}[\Delta\delta]}
     -
     \frac{\mathcal{R}_{2-3}}{F^{\mathcal{R};2-3}[\Delta\delta]}\right)\nonumber\\
	  &\times\frac{F^{\mathcal{R};1-2}F^{\mathcal{R};2-3}}
     {F^{\mathcal{R};1-2}[\Delta\delta]K^{\mathcal{R};2-3}
     -F^{\mathcal{R};2-3}[\Delta\delta]K^{\mathcal{R};1-2}}.
     \label{eq:mfdm_A4r}
\end{align}

The modified finite difference method can remove the $\mathcal{O}(\Delta\delta^2)$ terms coming from higher-order log terms. Since these terms are removed, the precision of the extracted values of 
\{$A_{4}$, $A_{4;l}$, $A_{6}$, $A_{6;l}$\} is increased. 
We will show below that these relations can be used to obtain stable values of 
\{$A_{4}$, $A_{4;l}$, $A_{6}$, $A_{6;l}$\} by tuning the uniform grid spacing $\Delta\delta$. The constants used to extract $\{A_4, A_{4;l}\}$ are listed in Table \ref{table:MFDM}.

\section{Numerical results}
\label{results}

In this section, we employ the strategy outlined above to extract the numerical values for the high-order isospin-asymmetry coefficients in the nuclear equation of state. We consider contributions from two-body forces and three-body forces up to second order in many-body perturbation theory. For three body forces at second order, we employ an in-medium NN interaction $V_{NN}^{\rm med}$ \cite{holt09,holt10,hebeler10} derived from the N2LO chiral 3N force, neglecting the explicit dependence on the center-of-mass momentum that arises due to the presence of the medium. We show that the normal finite difference method works well for both NN and 3N forces at the Hartree-Fock level. Comparing our numerical results using $V_{NN}^{\rm med}$ with exact calculations in Ref.\ \cite{Kaiser2012}, we find very good agreement despite the approximations that enter into the derivation of $V_{NN}^{\rm med}$. We then consider the equation of state at second-order in perturbation theory and use the modified finite difference method to extract the coefficients of the regular terms \{$A_4, A_6$\} and the logarithmic terms \{$A_{4;l}, A_{6;l}$\} in Eq.~\eqref{logs} when different potentials from chiral EFT are applied. We also study the density dependence of these higher-order isospin-asymmetry coefficients.

\subsection{NN and 3N forces in Hartree-Fock approximation}

The Maclaurin series coefficients of the isospin asymmetry expansion of the nuclear equation of state from both two-body and three-body chiral nuclear forces have already been studied in Ref.\ \cite{wellenhofer16} at the Hartree-Fock approximation. Here we review and verify some of the key results from Ref.\ \cite{wellenhofer16} and expand the discussion related to three-body forces. In all cases we find that the Hartree-Fock approximation admits a Maclaurin series expansion in even powers of the isospin asymmetry $\delta$.

\begin{figure}[t]
     \centering
     \includegraphics[scale=0.55]{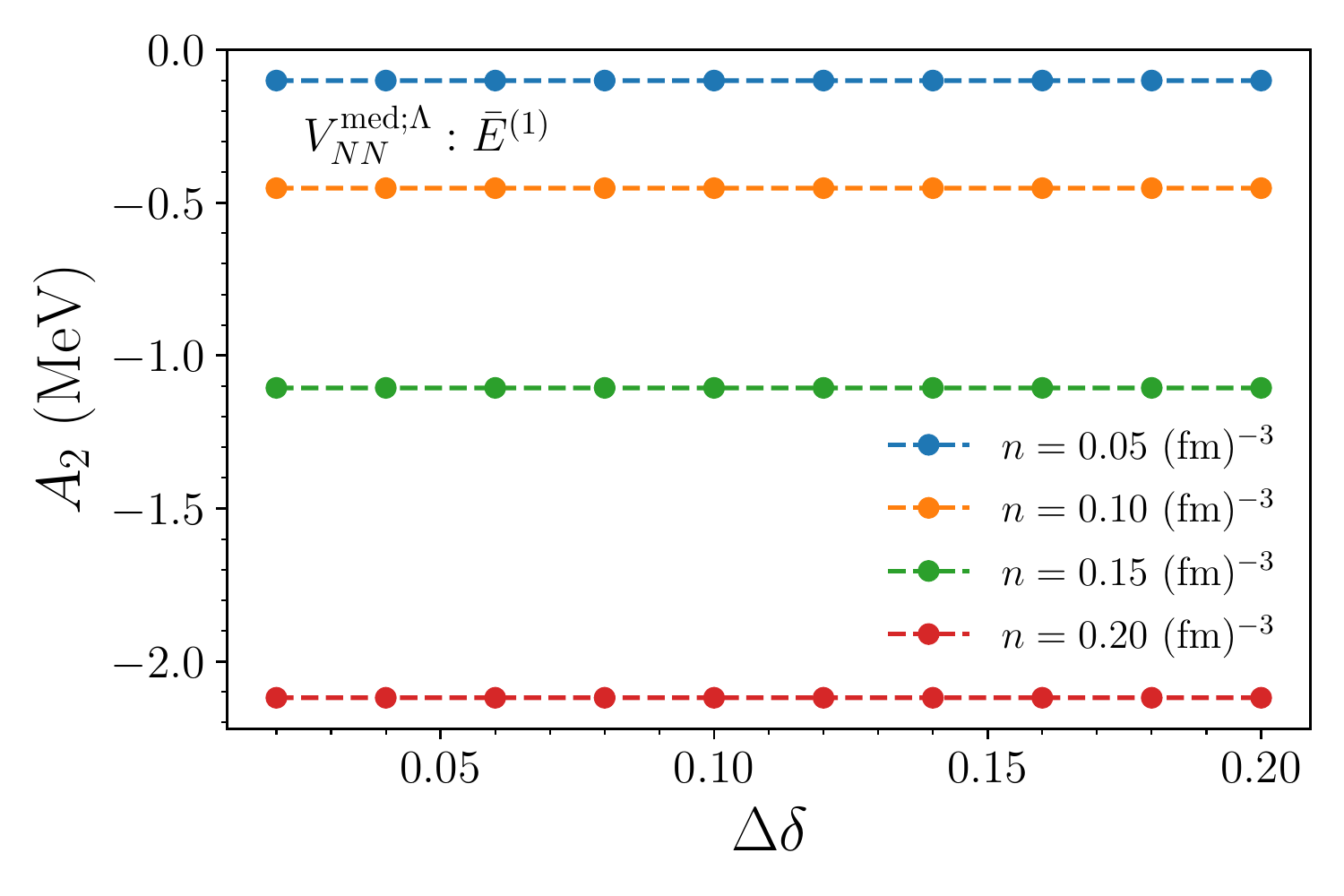}
	  \caption{Numerical values of the $A_2$ Maclaurin series coefficient from the N2LO chiral 3NF with low-energy constants $c_E = -0.106$, $c_D=-0.24$, $c_1=-0.81\rm{GeV}^{-1}$, $c_3=-3.4\rm{GeV}^{-1}$ and $c_4=3.4\rm{GeV}^{-1}$ at different densities and in the Hartree-Fock approximation. The finite difference method step sizes $\Delta \delta$ are varied, and the accuracy is chosen to be $\mathcal{A}=3$.}
     \label{fig:A2_sta_3NF_HF}
\end{figure}

In Fig.\ \ref{fig:As_2NF_HF} we show the density dependence of the Maclaurin series coefficients $A_2$, $A_4$, and $A_6$ when the nuclear EOS is calculated at the Hartree-Fock ($\bar E^{(1)}(\rho,\delta)$) level from five different N2LO and N3LO two-body potentials with cutoffs varying from $\Lambda=414, 450, 500$\,MeV \cite{entem03,marji13}. We observe that $A_2$ and $A_6$ grow monotonically with the density, but $A_4$ starts out positive at low densities and has a negative second derivative that in most cases causes the sign of $A_4$ to change. In all cases, the trends are consistent across the five different chiral potentials considered. Since the fourth-order and sixth-order terms often contribute with opposite sign, the dominant role played by $A_2$ is enhanced. Although there is clear evidence that $A_2 \gg A_4$ and $A_2 \gg A_6$, we find that $A_4$ and $A_6$ are comparable in the density range $0 < n < 0.20$\,fm$^{-3}$. However, in Fig.\ \ref{fig:A8_2NF_HF} we show the finite difference method extraction of the $A_8$ Maclaurin series coefficient for various chiral NN interactions at the density $n=0.15$\,fm$^{-3}$ for various spacings $\Delta \delta$. The plateaus in $\Delta \delta$ indicate that the value of $A_8$ is much less than $A_4$ and $A_6$, which suggests that the Maclaurin series converges efficiently. Although the behavior shown in Fig.\ \ref{fig:A8_2NF_HF} is just for a single density, we have verified that the coefficient remains small at other densities.

\begin{figure}[t]
     \centering
          \includegraphics[scale=0.42]{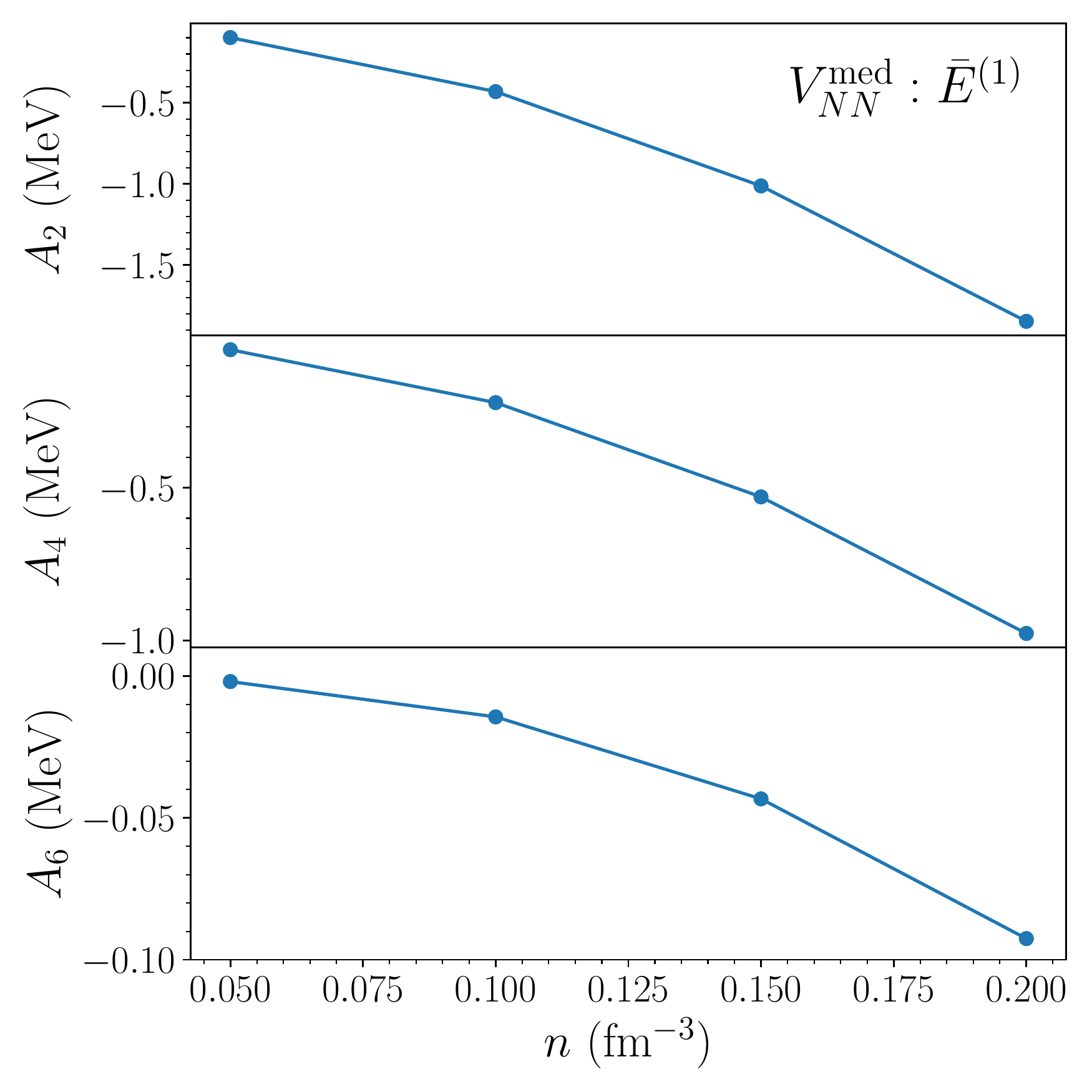}
     \caption{Density dependence of the Maclaurin series coefficients $A_2$, $A_4$, and $A_6$ of the isospin asymmetry expansion of the nuclear EOS at Hartree-Fock level from the N2LO 3-body interaction with low-energy constants $c_E = -0.106$, $c_D=-0.24$, $c_1=-0.81\rm{GeV}^{-1}$, $c_3=-3.4\rm{GeV}^{-1}$ and $c_4=3.4\rm{GeV}^{-1}$.}
     \label{fig:A246_3NF_HF}
\end{figure}

\begin{figure*}[t]
     \centering
     \includegraphics[scale=0.40]{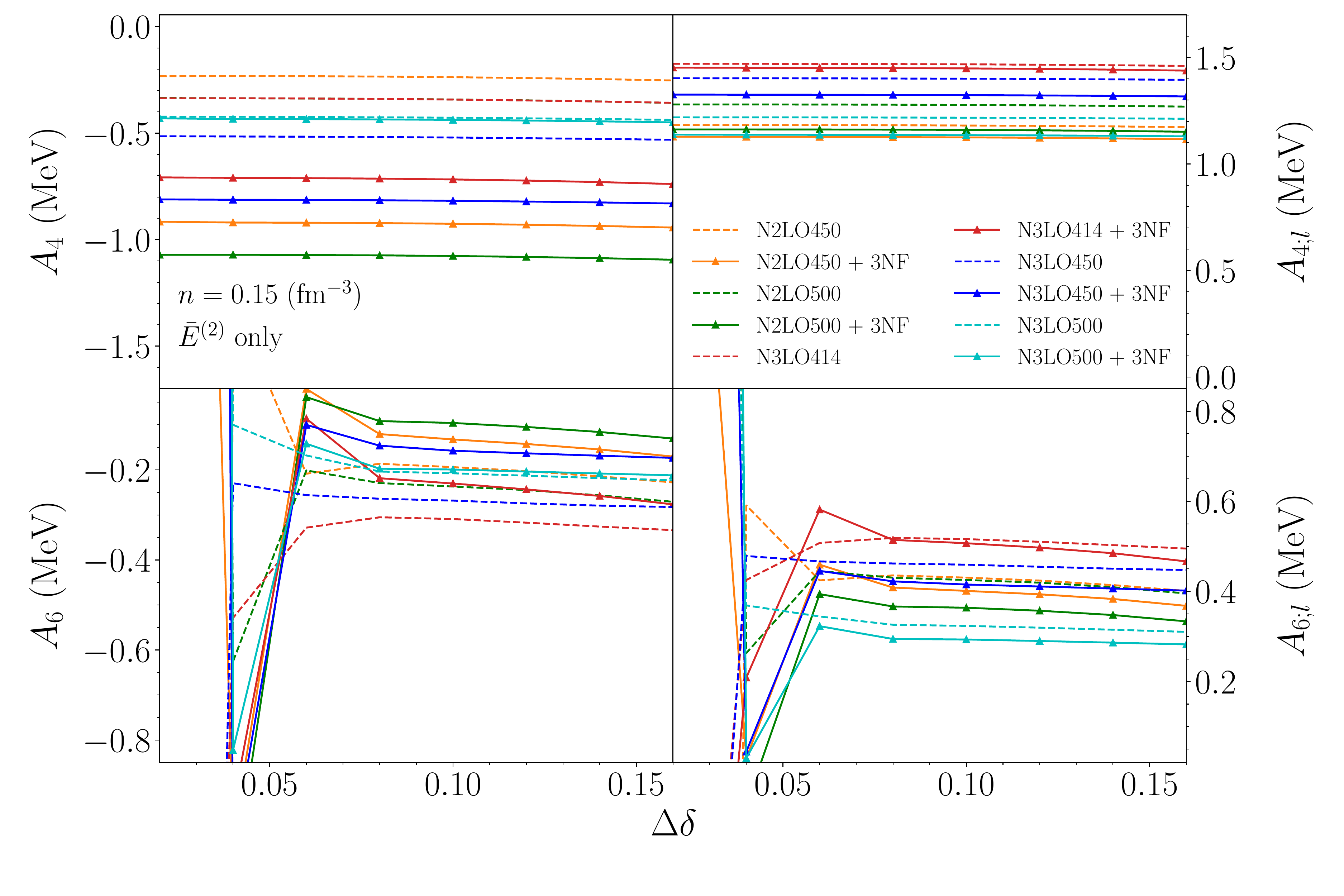}
	  \caption{Coefficients of the $4^{\rm th}$ and $6^{\rm th}$-order terms in $\delta$ for the isospin asymmetry expansion of the nuclear EOS from the second-order perturbation theory contribution $\bar E^{(2)}$. The nuclear density is fixed at $n=0.15$\,fm$^{-3}$, and we show results from five different chiral nuclear potentials with three-body forces (solid symbols) and without (dashed lines).}
     \label{fig:Arl_E2_pot}
\end{figure*}

We next compare the $A_2$ coefficients from the N2LO chiral 3NF obtained without any approximating assumptions in Ref.\ \cite{Kaiser2012} to those from the density-dependent NN interaction $V_{NN}^{\rm med}$ calculated in this work. Specifically, we build the in-medium NN interaction \cite{holt20} by summing the third particle over the filled states $\alpha$ in the noninteracting Fermi sea:
\begin{align}
	\hat V_{NN}^{\rm med} = \frac{1}{4}\sum_{\alpha ij lm}
	\langle \alpha ij| V_{3N}| \alpha lm\rangle 
	\hat{a}_i^\dagger \hat{a}_j^\dagger\hat{a}_m \hat{a}_l,
\end{align}
where
\begin{align}
	\sum_\alpha \to \sum_{s_\alpha t_\alpha}\int \frac{d^3 k_\alpha}{(2\pi)^3}\theta(k_f - k_\alpha).
\end{align}
In general, the medium will induce contributions to $V_{NN}^{\rm med}$ that depend on the center-of-mass momentum $\vec P$. In the following we construct $V_{NN}^{\rm med}$ by setting $\vec P=0$ and in addition assuming on-shell scattering for which the incoming and outgoing two-particle relative momenta are equal: $|\vec p\, | = |\vec p^{\,\prime}|$. These approximations result in a medium-dependent NN interaction with the same form as the free-space NN potential.

The N2LO chiral three-body interaction consists of a contact interaction $V^{ct}_{3N}$ proportional to the low-energy constant $c_E$, a one-pion-exchange interaction $V^{1\pi}_{3N}$ proportional to the low-energy constant $c_D$, and a two-pion-exchange interaction $V^{2\pi}_{3N}$ with terms proportional to the low-energy constants $c_1, c_3, c_4$. In Fig.\ \ref{fig:A2_3NF_HF}, we show as the solid blue curves the exact $A_2$ coefficients for $V^{ct}_{3N}$, $V^{1\pi}_{3N}$, and $V^{2\pi}_{3N}$ as a function of density without any applied regulator or simplifying assumptions \cite{Kaiser2012}. The low-energy constants associated with these terms of the chiral N2LO interaction are $c_E = -0.106$, $c_D=-0.24$, $c_1=-0.81\rm{GeV}^{-1}$, $c_3=-3.4\rm{GeV}^{-1}$ and $c_4=3.4\rm{GeV}^{-1}$. We also show as the orange dots the results from the associated density-dependent NN interaction $V_{NN}^{\rm med}$ without a high-momentum regulator. In both cases the coefficient $A_2$ is extracted 
by the normal finite difference method with accuracy $\mathcal{A}=3$. From the good agreement between these two calculations, we find that neglecting the explicit center-of-mass dependence when constructing $V_{NN}^{\rm med}$ does not lead to any noticeable modification of the leading $A_2$ isospin-asymmetry coefficients. In Fig.\ \ref{fig:A2_3NF_HF} we show with the green triangles in the lower-right panel the effect of the momentum-space regulating function
\begin{align}
     f(p',p) = e^{-(p'/\Lambda)^{2n}-(p/\Lambda)^{2n}},
\label{2Nreg}
\end{align}
where $\vec{p} = (\vec{p}_1-\vec{p}_2)/2$ and $\pvec{p}'=(\vec{p}_3-\vec{p}_4)/2$ are the relative momenta for the general two-body scattering process $N(\vec{p}_1)+(\vec{p}_2)\to N(\vec{p}_3)+N(\vec{p}_4)$ and $\Lambda = 450$\,MeV. We use the value $n=3$ so that the regulator implicitly affects only high powers in the chiral expansion \cite{holt20}. Although there are several different options for the choice of regulator \cite{holt20}, Eq.\ \eqref{2Nreg} is consistent with the regulator typically imposed on the free-space NN interaction and has relatively small artifacts \cite{holt20} compared to common local regulators. Naturally, the cutoff artifacts increase with density and reach about 10\% at $n=0.2$\,fm$^{-3}$ as shown in Fig.\ \ref{fig:A2_3NF_HF}. This cutoff dependence comes almost entirely from the two-pion-exchange contribution $V^{2\pi}_{3N}$, which consists of a series of strongly cancelling attractive and repulsive terms \cite{holt10}.

In Fig.~\ref{fig:A2_sta_3NF_HF} we demonstrate the stability of the $A_2$ extraction with respect to 
different grid spacings $\Delta\delta$ and different densities, using the same N2LO chiral potential with the cutoff $\Lambda=450$\,MeV analogous to the green triangles in Fig.\ \ref{fig:A2_3NF_HF}. The different colors \{blue, orange, green, red\} correspond to the densities $\{0.05, 0.10, 0.15, 0.20\}$\,fm$^{-3}$. We note that the normal finite difference method works extremely well for all densities considered. We also observe that the influence from higher-order terms almost vanishes at accuracy ${\cal A} = 3$ and therefore we can take the numerical value produced by the finite difference method as the true value of $A_{2}$ in Eq.~\eqref{eq:eos_exp_hf}. The 3NF contribution to the symmetry energy is rather small, on the order of $1-2$\,MeV at saturation density compared to the total empirical value of $A_2 \simeq 30$\,MeV, but it remains negative and increases approximately quadratically in magnitude with the density. We therefore see that at the Hartree-Fock approximation, the symmetry energy is dominated by two-body, rather than three-body, interactions.

\begin{figure}[t]
     \centering
      \includegraphics[scale=0.5]{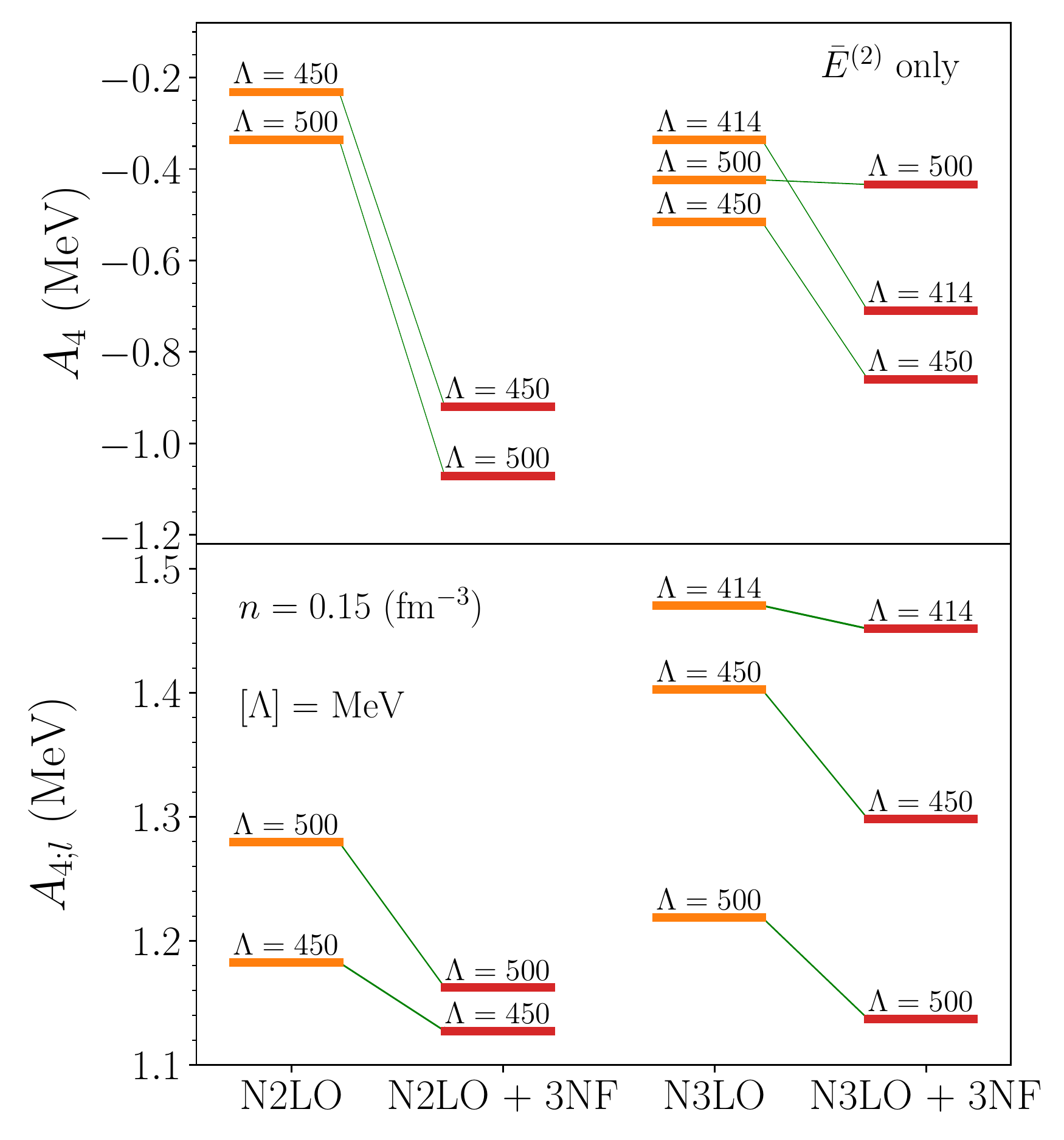}
	  \caption{The effect of the in-medium NN interaction $V_{NN}^{\rm med}$ derived from the N2LO chiral three-body force on the values of $A_4$ and $A_{4;l}$ for the second-order perturbative contribution $\bar E^{(2)}$.}
     \label{fig:A4rl_3NF}
\end{figure}

In Fig.\ \ref{fig:A246_3NF_HF} we show the density dependence of the Maclaurin series coefficients $A_2$, $A_4$, and $A_6$ in the isospin asymmetry expansion of the nuclear EOS at the Hartree-Fock level from the $\Lambda = 450$\,MeV chiral three-body force. We see that the higher-order symmetry energy coefficients systematically decrease in magnitude from one order to the next across all densities. We also observe that all Maclaurin series terms have roughly the same quadratic dependence on the density.

\subsection{Coefficients at $2^{\rm nd}$-order in perturbation theory}
In recent works \cite{kaiser15,wellenhofer16} it has been shown that the Maclaurin series form of the isospin asymmetry expansion of the nuclear EOS at second-order in perturbation theory is broken by divergent logarithmic terms that arise beyond the quadratic $\delta^2$ term in Eq.~\eqref{logs}. We will demonstrate that the modified finite difference method outlined in Section \ref{sec:fdiff} can be used to successfully extract the coefficients of both the regular and non-analytic logarithm terms in the expansion. In the following we calculate the second-order perturbation theory contribution $\bar E^{(2)}(\rho,\delta)$ and extract coefficients up to $\mathcal{O}(\delta^6)$. We employ several different nuclear two- and three-body potentials from chiral EFT by varying the order in the chiral expansion (N2LO and N3LO) and the momentum-space cutoff $\Lambda = 414, 450, 500$\,MeV. In all cases where three-body forces are included, we use the associated in-medium NN force $V_{NN}^{\rm med}$, which was shown in the preceding subsection to accurately capture the isospin-asymmetry dependence of the EOS at the Hartree-Fock level.

In Fig.\ \ref{fig:Arl_E2_pot} we show the results of the modified finite difference method for the extraction of the coefficients $\{A_4, A_{4;l}, A_6, A_{6;l}\}$ at a single value of the density $n=0.15$\,fm$^{-3}$ for five different chiral potentials. The dashed lines denote results for these coefficients when only two-body forces are included, while the solid lines also include the effects of three-body forces. Due to the influence from the high-order terms in $\delta$, the coefficients are not as stable with variations in $\Delta \delta$ as the Maclauin series coefficients derived at the Hartree-Fock level in the previous subsection, but overall we find that the high-order coefficients can be extracted very well from the modified finite difference method. At low values of $\Delta \delta$, the coefficients are difficult to extract due to the very small contribution to $\bar E^{(2)}$ from the high-order expansion terms. At large uniform grid spacing $\Delta\delta$, especially for the coefficients of the $6^{\rm th}$-order terms in $\delta$, numerical uncertainties arise from two sources. The first is due to the nature of the derivative, namely, the higher the order of the derivative, the larger the prefactor dropped from the index of the power. The second source of error comes from the modified finite difference method, which needs more points when it is used to extract coefficients of a higher-order term of $\delta$. This means that the value of $m$ in Eq.~\eqref{eq:MFDM} will be larger. The constants in Eq.~\eqref{eq:mfdm_A4l} and Eq.~\eqref{eq:mfdm_A4r} that absorb $\log(\sqrt{m})$ also become larger and enhance the influence from higher-order terms of $\delta$. Thus the numerical results from the modified finite difference method become more sensitive to higher orders. 

\begin{figure}[t]
     \centering
     \includegraphics[scale=0.55]{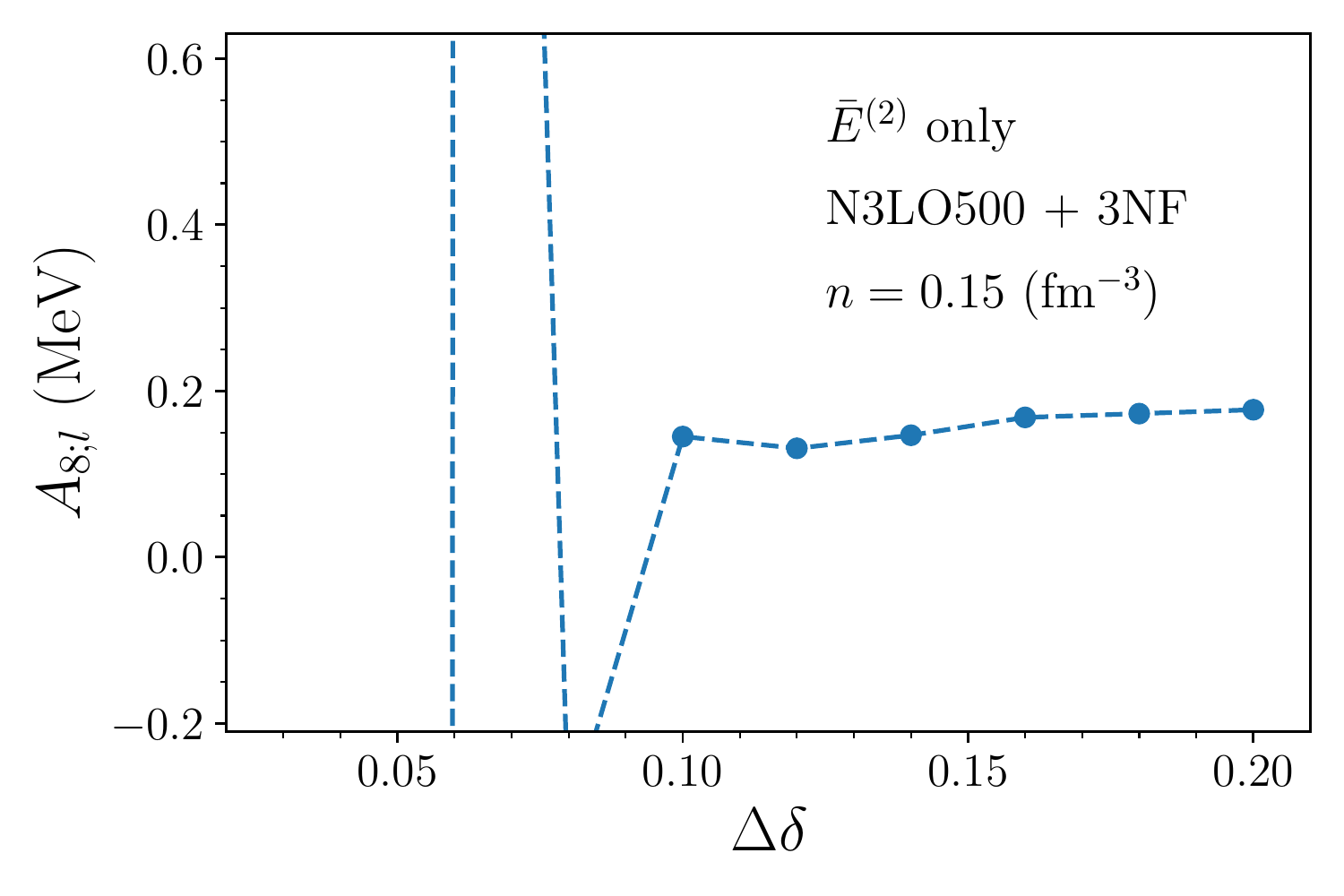}
	  \caption{Coefficient $A_{8;l}$ of the eighth-order logarithmic term in the modified isospin-asymmetry expansion of the nuclear EOS at second-order in perturbation theory. The density is chosen to be $0.15$\,fm$^{-3}$ and the potential is chosen to be the N3LO500 nuclear potential, including three-body forces.}
	  \label{fig:A8log_E2}
\end{figure}

\begin{figure*}[t]
     \centering
     \includegraphics[scale=0.45]{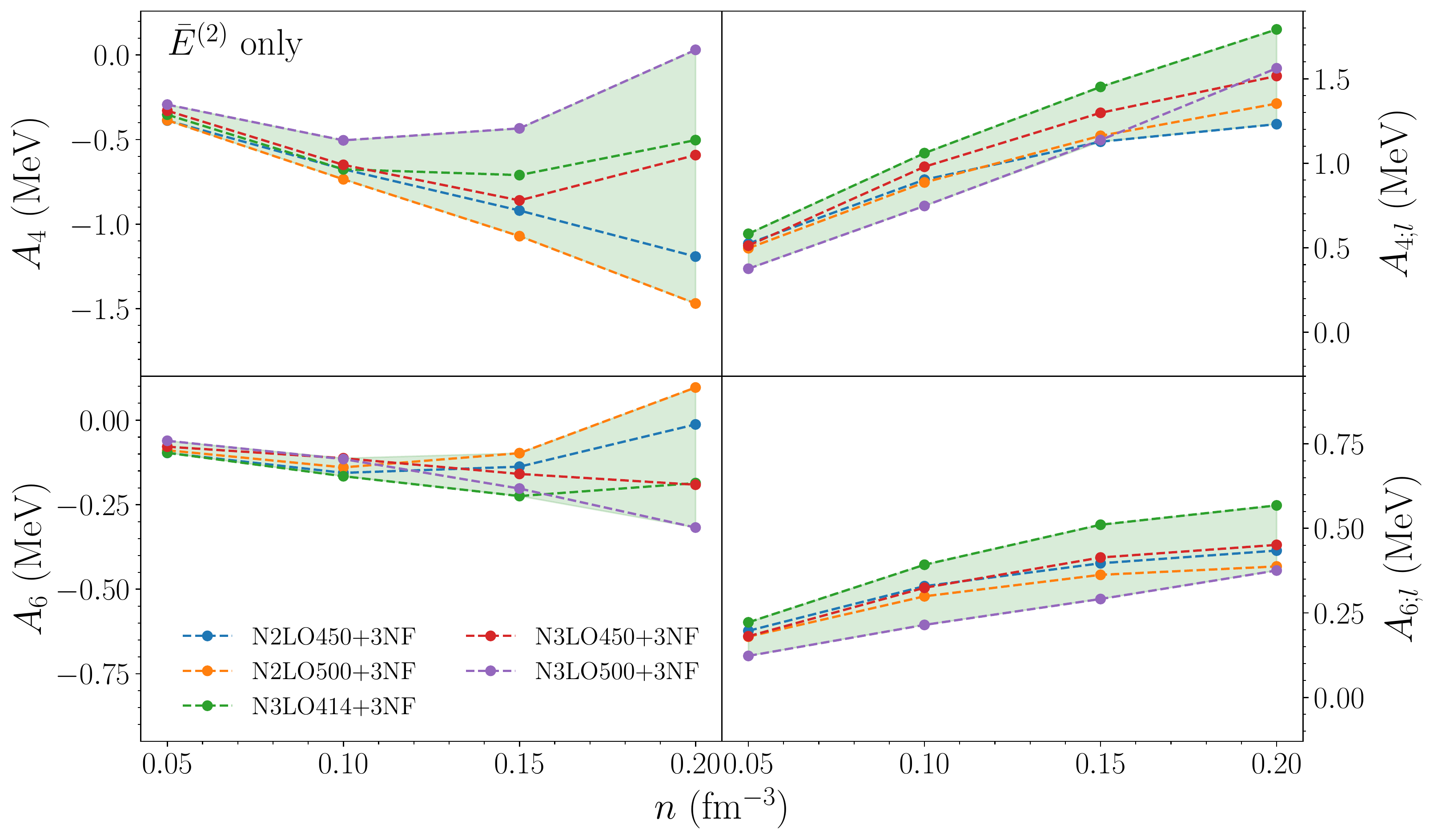}
	  \caption{Density dependence of the coefficients for the regular and nonanalytic terms in the isospin-asymmetry expansion of the EOS for the second-order perturbation theory contribution $\bar E^{(2)}$. Five different chiral NN + 3N forces are considered. Due to small uncertainties in the extraction of the $6^{\rm th}$-order terms in $\delta$, the plotted values of $A_{6}$ and $A_{6;l}$ are computed as the mean values among stable points within suitable regions of $\Delta\delta$ (compare Fig.\ \ref{fig:Arl_E2_pot}).}
     \label{fig:A46rho_E2}
\end{figure*}

The stability of these coefficients over some suitable region of $\Delta\delta$ can be found for all of the different chiral potentials, though some exhibit more stable plateaus than others. This indicates that the existence of $\log$ terms is not the property of a specific potential, but rather the property of second-order perturbation theory. Since the modified finite difference method is strongly dependent on the form of the expansion, the stability also suggests that there are no other types of nonanalytic terms. The relative uncertainty in the fourth- and sixth-order coefficients with regard to the choice of potential is smaller for the logarithmic contributions. Specifically, we find the ranges
\vspace{.2in}
\begin{align}
-1.07\,{\rm MeV} &< A_4 < -0.43\,{\rm MeV}, \nonumber \\
1.13\,{\rm MeV} &< A_{4;l} < 1.45\,{\rm MeV}, \nonumber \\
-0.28\,{\rm MeV} &< A_6 < -0.10\,{\rm MeV}, \nonumber \\
0.29\,{\rm MeV} &< A_{6;l} < 0.46\,{\rm MeV}. \nonumber
\end{align}
From Fig.\ \ref{fig:Arl_E2_pot} we see that at a given order of $\delta$, the magnitude of the logarithmic contribution ($A_{2i;l}$) is generically larger than that of the regular term ($A_{2i}$), a trend that persists across a wide range of densities as we will see below.

The effect of 3-body forces on the fourth-order terms in the isospin-asymmetry expansion of the EOS are shown in Fig.\ \ref{fig:A4rl_3NF} for all five chiral potentials considered in the present work. Results for only a single value of the density ($n=0.15$\,fm$^{-3}$) are shown. We see that three-body forces tend to strongly enhance the fourth-order regular term proportional to $\delta^4$ and reduce the strength of the fourth-order logarithmic contribution proportional to $\delta^4 \log|\delta|$. These effects are driven primarily by the two-pion exchange contribution $V_{3N}^{2\pi}$ to the N2LO chiral three-body force.

To better understand the convergence of the logarithmic terms in the modified expansion of the isospin asymmetry energy, we show in Fig.\ \ref{fig:A8log_E2} the $A_{8;l}$ coefficient from the modified finite difference method as a function of the spacing $\Delta \delta$. As a representative example, we consider the N3LO500 chiral nuclear potential at density $n=0.15$\,fm$^{-3}$. Comparing the values of $A_{4;l} \simeq 1.14$\,MeV, $A_{6;l} \simeq 0.29$\,MeV, and $A_{8;l} \simeq 0.18$\,MeV, we see that the present trend suggests a converging series for values of $\delta$ in the physical range $0<\delta<1$. As the order of the coefficient increases, we naturally find that it is more difficult to numerically extract the value from the modified finite difference method. Especially at low values of $\Delta \delta$, from Figs.\ \ref{fig:A4rl_3NF} and \ref{fig:A8log_E2} we see that numerical instabilities arise at $\Delta \delta_4 < 0.02$, $\Delta \delta_6 < 0.06$, and $\Delta \delta_8 < 0.1$ respectively. Even though the numerical value of the coefficient $A_{8;l}$ at such a high order in $\delta$ is quite sensitive to the numerical precision and higher-order terms of $\delta$, 
which lead to the divergence when $\Delta\delta$ is small and numerical noise when $\Delta\delta$ is large, respectively, one can still speculate a meaningful plateau in Fig.\ \ref{fig:A8log_E2}. For other chiral interactions, the plateaus are in general more challenging to isolate for this $A_{8;l}$ term.

In Fig.\ \ref{fig:A46rho_E2} we show the density dependence of the coefficients $\{A_4, A_{4;l}, A_6, A_{6;l}\}$ in the modified isospin-asymmetry expansion of the nuclear EOS from only the second-order perturbation theory contribution $\bar E^{(2)}$. We consider as an estimate of the theoretical uncertainty five different chiral potentials and include three-body forces throughout. At low values of the density, we see that all chiral potentials give quite similar results for all of the high-order isospin asymmetry coefficients. As the density increases past saturation density $n_0$, however, we find that the uncertainties in the normal terms $A_4$ and $A_6$ increase more significantly than the logarithmic terms. Especially the uncertainty in the $A_4$ term, which gives the sub-dominant contribution to the isospin-asymmetry energy after $A_2$, is large just beyond saturation density. Although the coefficients generically satisfy $|A_4| < |A_{4;l}|$ and $|A_6| < |A_{6;l}|$, for values of $\delta > 1/e \simeq 0.368$ the multiplicative factor $\delta^{2i}$ becomes larger than $\delta^{2i} \log|\delta|$ and therefore the impact of the nonanalytic logarithm terms on the isospin asymmetry expansion is weakened.

After extracting the coefficients $\{A_2$, $A_4$, $A_{4;l}$, $A_6$, $A_{6;l}$\}, we are able to expand the EOS up to different orders of $\delta$ and include nonanalytic terms in Eq.~\eqref{logs}. The difference $\Delta\bar{E}$ between the exact value of $\bar E^{(2)}(n,\delta)$ and the approximate value given by the sum of the isospin asymmetry expansion terms can be reduced substantially when high-order regular and nonanalytic $\log$ terms are included. We compare the precision of the widely-used approximation $\bar{E}_2 \equiv A_0 + A_2 \delta^2$ and the following approximations for the second-order perturbation theory contribution $\bar E^{(2)}$ to the EOS:
\begin{align}
	\bar{E}_{2i}^{(2)}(n,\delta) =& \bar{E}_2^{(2)} 
	+ \sum_{m\ge 2}^{2(i-1)}(A_{2m}^{(2)}+A_{2m;l}^{(2)}\log|\delta|)\delta^{2m}\nonumber\\
	&+ A_{2i}\delta^{2i},\label{eq:E2ir}\\
	\bar{E}_{2i;l}^{(2)}(n,\delta) =& \bar{E}_2^{(2)}
	+ \sum_{m\ge 2}^{2i}(A_{2m}^{(2)}+A_{2m;l}^{(2)}\log|\delta|)\delta^{2m}.\label{eq:E2il}
\end{align}

\begin{figure}[t]
     \centering
          \includegraphics[scale=0.56]{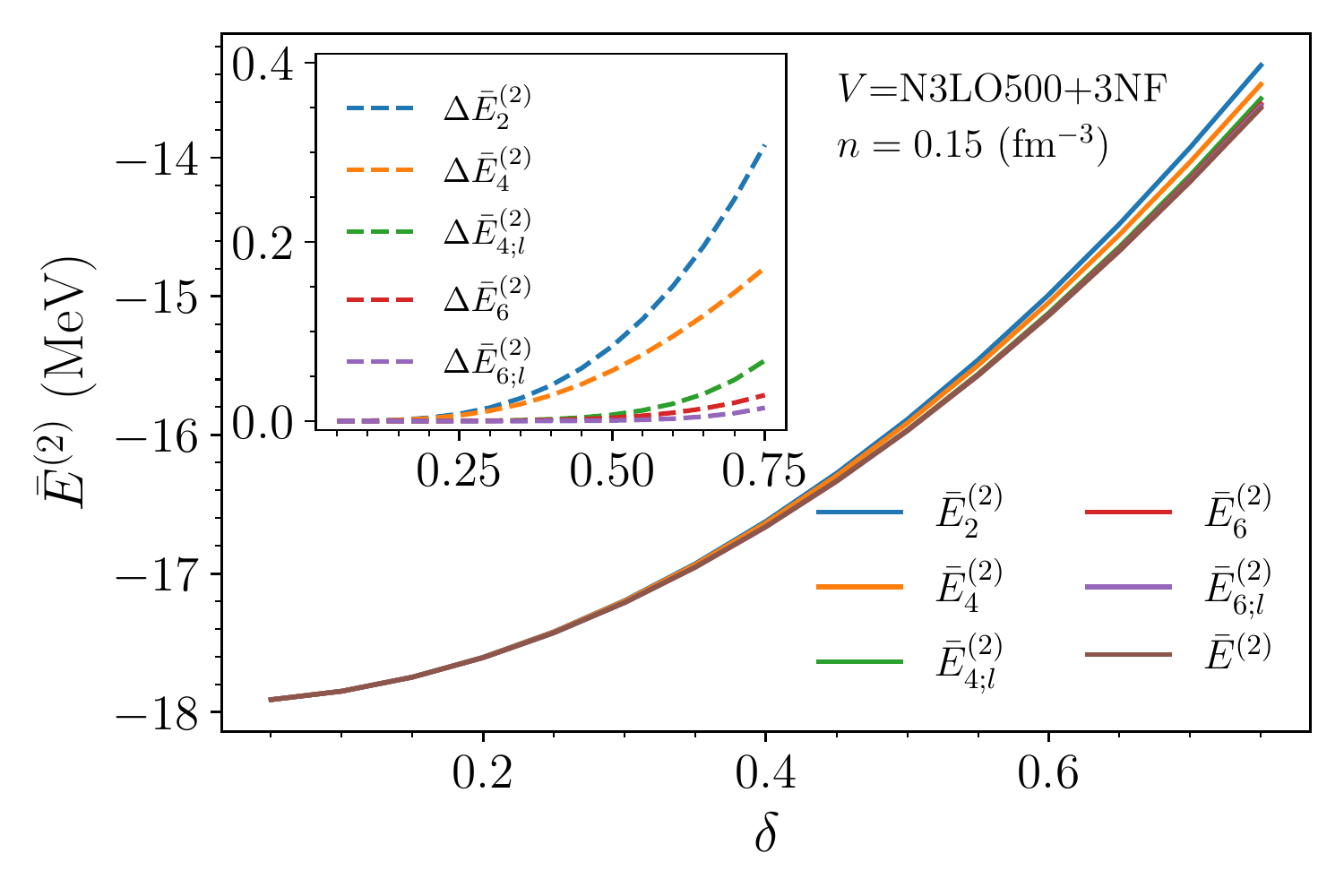}
	  \caption{Comparison between the approximate values of $\bar E^{(2)}$ given by Eqs.~\eqref{eq:E2ir} and \eqref{eq:E2il} 
	  with the exact value for fixed density $n=0.15$\,fm$^{-3}$ and varying isospin asymmetry $\delta$ from the N3LO500 chiral NN + 3N potential. The inset shows the numerical difference between the approximate values and the exact value, e.g., $\Delta\bar{E}_{4}^{(2)}=\bar{E}_{4}^{(2)}-\bar E^{(2)}$.}
     \label{fig:dif_comp}
\end{figure}

In particular, the $\bar{E}_{2i}^{(2)}$ terms do not include the associated $\log$ terms that arise at the same order. The comparison between $\bar{E}_{2i}^{(2)}$ and $\bar{E}_{2i;l}^{(2)}$ can give insight into the importance of the nonanalytic term in the isospin-asymmetry expansion of the EOS. In Fig.~\ref{fig:dif_comp}, we compare the exact value of $\bar E^{(2)}(\delta)$ at the fixed density $n=0.15$\,fm$^{-3}$ with the approximate values given by Eqs.~\eqref{eq:E2ir} and \eqref{eq:E2il}. We see that the inclusion of the high-order regular terms in even powers of $\delta$ significantly improves the precision, but the inclusion of the $\log$ terms is necessary for high precision, especially in the neutron-rich region where the functions $\delta^{2i} \log|\delta|$ peak.

\section{Proton fraction in $\beta$-equilibrium nuclear matter}
One motivation for the inclusion of higher-order terms in the isospin-asymmetry expansion of the nuclear EOS in Eq.~\eqref{logs} is to better understand the proton fraction in $\beta$-equilibrium nuclear matter found in neutron stars. In particular, the direct URCA process in nuclear matter is composed of the neutron decay and electron capture reactions shown in Eq.~\eqref{eq:beta_proc}. These two competing processes balance when the system satisfies the chemical equilibrium condition:
\begin{align}
	\mu_e = \mu_n - \mu_p,\label{eq:chemical_eq}
\end{align}
where $\mu_e$, $\mu_n$ and $\mu_p$ are the chemical potentials of electrons, neutrons, and protons respectively. The right-hand side of Eq.~\eqref{eq:chemical_eq} can be evaluated from the fundamental thermodynamic relation
\begin{align}
	\mu_n - \mu_p = -\frac{\partial \bar{E}}{\partial x} 
	= 2\frac{\partial \bar{E}}{\partial{\delta}},\label{eq:che_dif}
\end{align}
where $x = n_p/ n$ is the proton fraction. 
Since neutron stars rapidly cool to nearly degenerate conditions, we assume the temperature to be zero, in which case the chemical potentials of all species are just their respective Fermi energies. Since electrons are ultra-relativistic, $\mu_e = k^F_e$. The final condition to be imposed is charge neutrality: $n_p = n_e$. The combination of chemical equilibrium and charge neutrality determines the proton fraction $x=n_p/n$ in $\beta$-equilibrium nuclear matter.

\begin{figure*}[t]
	\centering
	\includegraphics[scale=0.45]{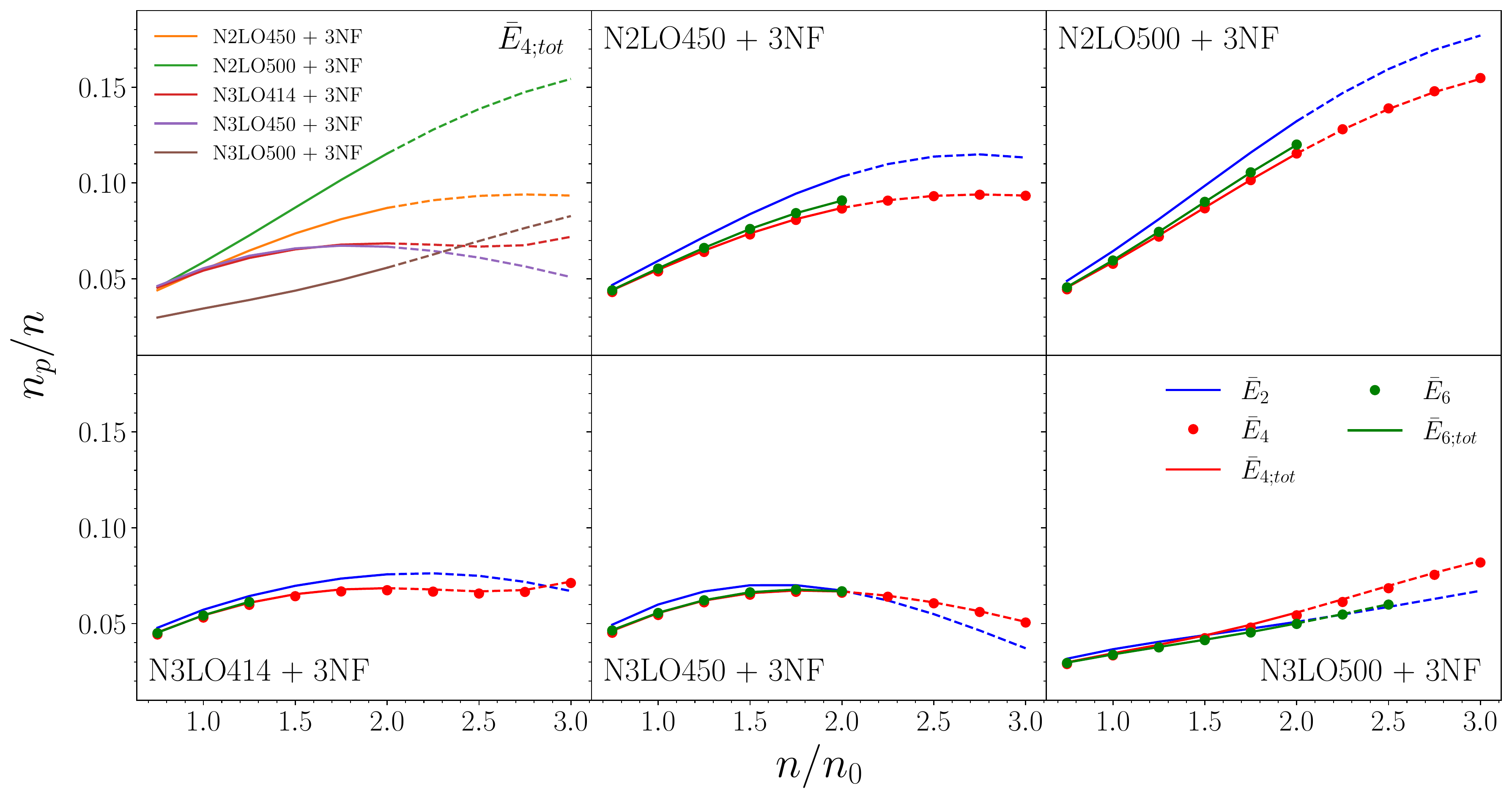}
	\caption{The proton fraction in $\beta$-equilibrium nuclear matter as a function of density $n$ for five different chiral NN + 3N potentials including different orders in the isospin-asymmetry expansion of the EOS. Beyond $n>2n_0$ we denote the predictions with dotted lines to indicate their speculative nature beyond the regime of validity for chiral EFT.}
	\label{fig:pro_fra}
\end{figure*}

We use different approximations for the EOS of nuclear matter to study the proton fraction in beta-equilibrated neutron stars and the role of high-order terms beyond $A_2$. In Fig.\ \ref{fig:pro_fra} we show the resulting proton fraction $n_p/n$ calculated from the equation of state of nuclear matter at second order in perturbation theory as a function of the total density $n/n_0$ normalized to saturation density for five different chiral nuclear potentials. In all cases, the effects of three-body forces are included. In the top-left panel, we show the density dependence of the proton fraction at the $\bar{E}_{4;tot}$ approximation, which is the highest order that we can extract precisely up to high densities using all of the chiral interactions. We find that there is a quite large uncertainty $0.06 < n_p/n < 0.12$ in the proton fraction at $n=2n_0$. Beyond twice saturation density (where chiral effective field theory may become unreliable) we plot speculative values of the proton fraction for all five chiral interactions. We find that the uncertainties continue to grow and already at $3n_0$ the proton fraction falls in the large range $0.05 < n_p/n < 0.15$.

In the remaining subpanels of Fig.\ \ref{fig:pro_fra} we show the individual effects of the high-order terms beyond $A_2$ on the proton fraction of $\beta$-equilibrium matter for each of the five chiral interactions. The changes due to fourth-order and sixth-order regular terms $A_{4}$ and $A_{6}$ are denoted with red and green filled circles, respectively, while the changes due to fourth-order and sixth-order logarithmic terms $\bar E_{4;l}$ and $\bar E_{6;l}$ are denoted with red and green solid lines. It is important to keep in mind that the coefficients of the regular terms $\bar E_{2i}$ come from three contributions: the noninteracting Fermi gas term, the Hartree-Fock term, and the second-order perturbation theory term, while $\bar E_{2i;l}$ only arises from the second-order perturbation theory term. We see that in general the logarithm terms do not have a large impact on the proton fraction. The differences between the proton fraction computed in the $\bar{E}_2$ approximation and in the $\bar{E}_{4;tot}$ approximation are relatively small around saturation density but grow to a relative strength around 10-20\% at $n=2n_0$. The sixth-order terms give a relatively small change to the proton fraction, but in the case of the N2LO chiral interactions the effects are non-negligible at high densities.

\section{summary}

We have examined the isospin-asymmetry expansion of the nuclear matter EOS from many-body perturbation theory using several high-precision NN and 3N chiral nuclear potentials. From a series of precise calculations of the ground-state energy of isospin-asymmetric nuclear matter at zero temperature, we have extracted normal and logarithmic contributions to the isospin-asymmetry expansion up to sixth order in the asymmetry parameter $\delta$. The equation of state calculated in the Hartree-Fock approximation contains no logarithmic terms, and the standard finite difference method can be used to extract the coefficients of the isospin-asymmetry expansion. By comparing the analytical results for the $A_2$ coefficient from the N2LO chiral 3NF in Ref.\ \cite{Kaiser2012} at the Hartree-Fock appoximation with the numerical extraction from the in-medium NN interaction $V_{NN}^{\rm med}$, we have shown that $V_{NN}^{\rm med}$ accurately reflects the isospin asymmetry dependence of the EOS induced by three-body forces. For the second-order perturbation theory contribution including NN and 3N forces, the stable extraction of both the normal and logarithmic terms could only be achieved with a modified finite difference method developed in the present work.

For all chiral nuclear potentials and densities considered in the present work, we have found that the modified isospin-asymmetry expansion converges relatively quickly. Although in some cases the fourth and sixth-order terms in the expansion of the isospin-asymmetry dependence of the EOS are of similar magnitude, all eighth-order terms appear to be an order of magnitude smaller. We have also found that at a given order of $\delta$, the coefficients $A_{2i;l}$ of the logarithmic terms are generically larger than the coefficients $A_{2i}$ of the regular terms in the $E^{(2)}$ contribution to the EOS. This feature is enhanced as the nuclear density increases. From the accurate numerical extraction of the high-order isospin asymmetry coefficients, we conjecture that no other nonanalytic terms beyond $\log |\delta|$ appear at second order in perturbation theory.

As an application of the above analysis, we have studied the proton fraction in $\beta$-equilibrium nuclear matter based on the approximate EOS obtained from the isospin asymmetry expansion. Overall, there is a large dependence of the proton fraction on the choice of chiral potential. Effects from fourth-order terms in the isospin-asymmetry expansion are important and in some cases can change the proton fraction by 2\% already at twice saturation density $2n_0$. Even though the magnitude of the coefficient of the $\log$ term is greater than of the regular term at a given order of $\delta$ at second-order perturbative level, the regular terms still dominate the proton fraction. We also find that the effects from sixth-order contributions to the isospin-asymmetry dependence of the EOS are small, but in some cases they can change the proton fraction by 0.5\%.

\bibliographystyle{apsrev4-1}
%

\end{document}